\documentclass[preprint,aps]{revtex4}

\usepackage{graphicx}
\usepackage{dcolumn}
\usepackage{bm}

\newcounter{sub}
\newcounter{subeqn}[sub]
\renewcommand{\thesubeqn}{\alph{subeqn}}
\renewcommand{\theequation}{\thesub\thesubeqn}

\def\be{\begin{equation}}
\def\ee{\end{equation}}

\def\ms{\left[}
\def\rs{\right]}

\def\st{\stepcounter{sub}}

\def\bea{\begin{eqnarray}}
\def\eea{\end{eqnarray}}

\newcommand\xxi{{\mbox{\boldmath $\xi$}}}

\newcommand\xis{{\mbox{\scriptsize{$\xxi$}}}}
\newcommand\mmu{{\mbox{\boldmath $\mu$}}}
\newcommand\llambda{{\mbox{\boldmath $\lambda$}}}

\newcommand\eeta{{\mbox{\boldmath $\eta$}}}
\newcommand\etas{{\mbox{\scriptsize{$\eeta$}}}}
\newcommand\nab{\mbox{\boldmath $\nabla$}}

\newcommand\psip{{\mbox{${\psi^+}$}}}
\newcommand\chip{{\mbox{${\chi^+}$}}}
\newcommand\phip{{\mbox{${\phi^+}$}}}

\def\k{{\bf k}}
\def\l{{\bf l}}
\def\m{{\bf m}}

\def\r{{\bf r}}
\def\v{{\bf v}}

\newcommand\A{{\bf A}}
\newcommand\I{{\bf I}}

\newcommand\B{{\bf B}}

\newcommand\C{{\bf C}}

\newcommand\no{\nonumber}

\newcommand\GTP{{\rm GTP}}
\newcommand\GDP{{\rm GDP}}
\newcommand\Tu{{\rm T}}
\newcommand\MT{{\rm MT}}

\begin{document}


\title{Modeling polymerization of microtubules: A semi-classical nonlinear field theory approach
}

\author{Vahid Rezania$^{1,2}$ and Jack Tuszynski$^1$\footnote{Corresponding author: Division of Experimental Oncology, Cross Cancer Institute,
11560 University Avenue, Edmonton, AB T6G 1Z2
Canada, Tel: +1(780)432 8906, Fax: +1(780)432 8892; Email: jtus@phys.ualberta.ca } }
\address{1- Division of Experimental Oncology, Cross Cancer Institute
11560 University Avenue, Edmonton, AB T6G 1Z2
Canada \\
2 - Department of Physics, MacEwan College, Edmonton, AB T5J 2P2, Canada
}

\begin{abstract}
In this paper, for the first time, a three-dimensional treatment of
microtubules' polymerization is presented.
Starting from fundamental
biochemical reactions during microtubule's assembly and disassembly processes,
we systematically derive a nonlinear system of equations
that determines the dynamics of microtubules in three dimensions. We
found that the dynamics of a microtubule is mathematically expressed via a
cubic-quintic nonlinear Schr\"{o}dinger (NLS) equation. We show that
in 3D a vortex filament, a generic solution of the NLS equation,
exhibits linear growth/shrinkage in time as well as temporal
fluctuations about some mean value which is qualitatively similar to
the dynamic instability of microtubules.
By solving equations numerically, we have found
spatio-temporal patterns consistent with experimental observations.
\end{abstract}


\maketitle

\section{Introduction}\label{int}
More than 25 years ago, Del Giudice \emph{et al.} \cite{Del85} argued that
a quantum field theory approach to the collective behaviors of biological
systems is not only applicable but the most adequate as it leads naturally to
nonlinear, emergent behavior which is characteristic of biological organization.
They followed a line of reasoning championed by Davydov \cite{Dav82,Dav79} and Fr\"ohlich \cite{Fro77}
who emphasized integration of both conservative and dissipative  mechanisms in biological
matter leading to the emergence of spatio-temporal coherence with various specific
manifestations such as almost lossless energy transport and long-range coordination.

Microtubules (MTs) are long protein polymers present in almost all
living cells. They participate in a host of cellular functions with
great specificity and spatio-temporal organization.  In most
multicellular organisms, the interior of each cell is spanned by a
dynamic network of molecular fibers called the \emph{cytoskeleton}
(`skeleton of the cell'). The cytoskeleton gives a cell its shape,
acts as a conveyor for molecular transport, and organizes the
segregation of chromosomes during cell division, amongst many other
activities. The complexity and specificity of its functions has
given rise to the notion that along with its structural and mechanical
roles, the cytoskeleton also acts as an information processor
\cite{Alb85}, or simply put the ``cell's nervous system"
\cite{Ham87}.   A microtubule is a hollow cylinder, a rolled-up
hexagonal array of tubulin dimers arranged in chains along the
cylinder (`protofilaments'). Within cells, microtubules come in
bundles held together by `microtubule associated proteins' (MAPs).
The geometry, behavior and exact constitution of microtubules varies
between cells and between species, but an especially stable form of
a microtubule runs down the interior of the axons of human neurons.
Conventional neuroscience at present ascribes no computational role
to them, but models exist in which they interact with the membrane's
action potential \cite{BT97,Pri06}.

Microtubules are very dynamic bio-polymers that simply lengthen and/or shorten
repeatedly at the macroscopic level during a course of time on the scale of minutes.
At the microscopic level, however, several biochemical reactions
are taking place in order for an individual microtubule to undergo an assembly or
disassembly process.
This dynamical behavior of microtubules (so-called dynamic instability)
has attracted many investigators for decades to examine microtubules'
behavior in many aspects (see section 2 for further details).

Although there is no systematic description for the microtubule's assembly/disassembly
process at the microscopic level, several theoretical models have been proposed to
describe the macroscopic or statistical aspect of lengthening/shortening of microtubules using nonlinear
classical equations \cite{Fey94,Bic97,BR99,Dog95,DY98,DL93}.
The latter studies while providing good agreements with experimental results, they are more or
less phenomenological.  Therefore, several features of the
microtubule's assembly/disassembly process might not be adequately captured.

In this paper, we propose a systematic model for the microtubule's
assembly/disassembly process at the microscopic level using a
first-principles quantum mechanical approach as a starting point. In this model we consider an individual
microtubule with length $L$ consisting of $N$ tubulin layers viewed here as a quantum state $|N\rangle$.
The state can be raised/lowered by a creation/annihilation operator
(i.e. polymerization/depolymerization process) to the $|N+1\rangle/|N-1\rangle$
state.  The corresponding microtubule is then longer/shorter by one tubulin
layer from the original one.
Based on the chemical binding reactions that are taking place during microtubule
polymerization, a quantum mechanical Hamiltonian for the system is proposed.
Equations of motion are then derived and transformed from the purely quantum mechanical
description to a semi-classical picture using the inverse Fourier  transformation. 
The resulting nonlinear field dynamics
reduces to the cubic-quintic nonlinear Schr\"{o}dinger equation that provides a richer dynamics
than the previous phenomenological descriptions and
includes both localized energy transfer and oscillatory solutions both of which have either been experimentally
demonstrated or theoretically predicted earlier.
We believe that this treatment is both useful and necessary to address the fundamental issues
about the observed dynamical behavior of MTs.  As stated by Del Giudice \emph{et al.} \cite{Del85} :
``Systems with collective modes are naturally described by field theories. Furthermore,
quantum theory has proven to be the only successful tool for describing
atoms, molecules and their interactions."

\section{Microtubule assembly background}
\noindent
A very rigid (by biological standards) and typically several micrometers long rod-like polymer plays an
essential role during cell division.    The so-called
microtubule (MT) is assembled by tubulin polymerization
in a helical lattice.  These protein polymers are
responsible for several fundamental cellular processes, such as
locomotion, morphogenesis, and reproduction \cite{ALRW94}.
It is also suggested that MTs are responsible for transferring energy
across the cell, with little or no dissipation.

Both \emph{in vivo} and \emph{in vitro} observations
confirmed that an individual MT switches stochastically
between assembling and disassembling states that makes MTs
highly dynamic structures \cite{MK84a,MK84b}.
This behavior of MTs is referred to as dynamic instability.
Dynamic instability of MTs is a nonequilibrium process that has
been the subject of extensive research for the past two decades.
It is generally believed that the instability starts from the hydrolysis
of guanosine triphosphate (GTP) tubulin that follows by converting GTP
to guanosine diphosphate (GDP).   This reaction is exothermic and
releases $\sim 8 k_{\rm B}T$ energy per reaction \cite{WIS89}, i.e. approximately
$0.22$ eV per molecule \cite{EAHH89}.  Here $k_{\rm B}$ is the Boltzmann's
constant and $T$ is the absolute temperature.   Since GDP-bound tubulin favors
dissociation, a MT enters the depolymerization phase as the advancing
hydrolysis reaches the growing end of a MT.   This phase transition is
called a catastrophe.  As a result of this transition, MTs start breaking
down, releasing the GDP-tubulin into the solution.  In the solution,
however, reverse hydrolysis takes place and a polymerization phase of
MTs begins utilizing assembly-competent tubulin dimers.  The latter phase transition which comes after a catastrophe
is called a rescue.   Therefore, MTs constantly fluctuate between growth
and shrinkage phases.

Odd \emph{et al.} \cite{OCB95} studied both experimentally and theoretically MT's assembly
to extract their catastrophe kinetics.
These authors  proposed that a growing MT
may remember its past phase states by analyzing growth characteristic of
both plus and minus ends of several individual MTs.   Their results
showed that while the minus end growth time follows an
exponential distribution, the plus end fits a gamma distribution.
The exponential (gamma) distribution suggests a
first (non-first) order transition between growing and shrinking
phases.  Statistically, the exponential distribution
represents that the new state happens independently of the
previous state.  As a result, a MT with first order catastrophe
kinetics does not remember for how long it has been growing.
  In contrast, the
catastrophe frequency of a MT with non-first order kinetics would
depend on its growth phase period.  The gamma distribution suggests
that the catastrophe frequency is close to zero at early times,
increases over time and reaches asymptotically a plateau.
This is consistent with observations that the
catastrophe events are more likely at longer times.
Odd \emph{et al.} \cite{OCB95} concluded that such behavior implies that a
`crude form of memory' may be built in MT's dynamic instability.
As a result, a microtubule would go through an `intermediate state'
before a catastrophe event takes place.

The dynamics of transitions between growing and shrinking states is still
a subject of controversy.  It is suggested that a growing MT has a
stabilizing cap of GTP tubulin at the end which keeps it from
disassembling \cite{MK84a,MK84b}.  Whenever MT loses its cap,
it will undergo the shrinking state.   Several theoretical and
experimental studies have been devoted to the cap model.  For the purpose of this paper,
we emphasize the link between GTP hydrolysis and the switching process
from growing to shrinking of a MT.   GTP hydrolysis is a subtle biochemical
process that carries a quantum of biological energy and thus allows us
to make a link between quantum mechanics and polymer dynamics.
We return to this theme later in the paper but first discuss the
statistical methods used in this area.

\subsection{Ensemble dynamics of microtubules}
\noindent
As we discussed earlier, the MT dynamical instability has been the subject of
numerous studies. Although the dynamical instability of MTs is a
nonlinear and stochastic process,
so far only their averaged behaviors have been analyzed using a simple model.  Introducing
$p_g(x,t)$ and $p_s(x,t)$ as the probability density of a growing and shrinking tip,
respectively, of a MT with length $x$ at time $t$, Dogterom and Leibler  \cite{DL93} proposed the following
equations for the time evolution of an individual MT:
\st
\bea \label{grow0}
&&\partial_t p_g = - f_{gs} p_g + f_{sg} p_s - v_g \partial_x p_g,\\
\st \label{shrink0}
&&\partial_t p_s =  f_{gs} p_g - f_{sg} p_s - v_s \partial_x p_s.
\eea
Here $f_{gs}$ and $f_{sg}$ are the transition rates from a growing to
a shrinking state and vice versa.  The average speeds of the MT in the assembly and
disassembly states are given by $v_g$ and $v_s$, respectively (see also
\cite{Bic97,BR99,DY98}).

Random fluctuations about the MT's tip location can be also modeled by adding a
diffusive term in the above equations:
\st
\bea \label{grow}
&&\partial_t p_g = - f_{gs} p_g + f_{sg} p_s - v_g \partial_x p_g + D_g\partial_{xx}p_g,\\
\st \label{shrink}
&&\partial_t p_s =  f_{gs} p_g - f_{sg} p_s - v_s \partial_x p_s + D_s\partial_{xx}p_s,
\eea
where $D_g$ and $D_s$ are the effective diffusion constants in the two states
\cite{FHL94,FHL96}.

Equations (\ref{grow}) and (\ref{shrink}) describe the overall dynamics of an individual MT
without considering the dynamics of GDP and GTP tubulin present in the solution.  It is
clear that the MTs are growing faster in the area with a higher concentration of GTP tubulin.
Using this fact, Dogterom \emph{et al.} \cite{Dog95} generalized the above model by incorporating the tubulin dynamics.
They added
two more equations to the above system:
\st
\bea \label{c_T}
&&\partial_t c_T = - v_g s_0  p_g + k c_D + D_1\nabla^2 c_T,\\
\st \label{c_D} &&\partial_t c_D =  v_s s_0  p_s - k c_D +
D_2\nabla^2 c_D,
\eea
where $c_T$ and $c_D$ are average
concentrations of GTP and GDP tubulin, respectively. $D_1$ and $D_2$
are the diffusion coefficients, $k$ is the rate constant and $0 \leq
s_0\leq 1$.   In view of the link to quantum transitions between GDP
and GTP at the root of this problem we now introduce a method that
allows a smooth transition from quantum to classical (nonlinear)
dynamics of MT assembly/disassembly process. Recently, Antal \emph{et al.}
\cite{Ant07_1,Ant07_2} also proposed a one-dimensional statistical
model to describe the dynamic instability of MTs.  They considered a
two state model that MT grows with a rate $\lambda$ and shrinks with a
rate $\mu$ and then obtained similar fluctuations in MT's length by
varying $\lambda$ and $\mu$ as observed for MTs.

The above mentioned studies generally considered a MT as
a one-dimensional mathematical object that grows and shrinks
with random rates.  Such a treatment is not only applicable to MTs but also
to a polymer system whose assembly or disassembly processes have an element of
randomness.  As a result, the biophysical and biochemical
characteristics of MTs cannot be adequately captured.
In this paper, however, we based our model on fundamental
biochemical reactions that are occurring during microtubule's assembly
and disassembly processes.  This allows us to derive a nonlinear system of equations
that determines the structure, the dynamics and the motion of MTs in 3D.

\subsection{Deriving Semiclassical Equations} 

The underlying method we use here 
has been developed in a number of papers and  a book
\cite{DT90b,DT90a,TD89a,TD89b,TD89c,TD89d,Tus90} and is essentially
semiclassical in nature. The treatment is quantitative in that important
terms which are retained are calculated exactly and those which are very
small but nevertheless significant are discussed at a later stage and
their effect is estimated.  The motivation for the method and a derivation
of the dynamical field equation are presented in \cite{TD89a} and a
discussion of the types of classical field solutions is presented in
\cite{TD89b}.  A fuller version has been published in the review paper
in \cite{Tus90} whereas a very brief overview is given in \cite{TD89c}.
It has been successfully applied to the phenomenon of superconductivity
\cite{TD89d,DT90a} and when combined with topological arguments
yields, for example, the correct temperature dependence of the critical
current density in low temperature superconductors.  One can also obtain
the position of phase boundaries in metamagnets where previously
only elaborate numerical techniques could provide this information
\cite{DT90b}.  Spatial correlations are fully incorporated using a
renormalization technique and quantum fluctuations have been
included also.
It has been demonstrated that even when the method is generalized
to include spin-dependent fields, the equation of motion for the field
is of the same form \cite{DT91} and the classical field equation is
also of the same form for both Boson and Fermion particles.  This does
not mean that the Fermionic character of the electrons disappears
because the statistics of the particles reappears in the
choice of the classical field which satisfies the physical boundary
conditions on the charge density.  The method is basically
non-relativistic although it could be readily generalized but here
we use the non-relativistic version.
The starting point in this method 
is to write a generic form of
second-quantized Hamiltonian using one particle state
annihilation and creation operators:
\st
\be
H=\sum_\k\hbar\omega_\k q_\k^\dagger q_\k
+ \sum_{\k,~\l,~\m} \hbar\Delta_{\k,~\l,~\m}q_\k^\dagger q_\l^\dagger
q_\m q_{\k+\l-\m},
\ee
where the vectors $\k,~\l$ and $\m$ are shorthand labels for
quantum numbers of a complete orthonormal set of particle functions
in the usual way and we use the linear momentum conserving form for the
two-body interaction.  Depending on the system studied,
using Fermi-Dirac or Bose-Einstein statistics one can derive the
Heisenberg's equation of motion:
\st
\be\label{mm}
i\hbar\partial_t q_\k(\r,t)=[H,q_\k(\r,t)].
\ee
Now both sides of Eq. (\ref{mm}) are multiplied by
$\Omega^{-1/2} \exp(-i \eeta\cdot\r) a_\etas (t)$ and summed over $\eeta$.
At the same time the matrix elements $\omega_\k$ and $\Delta_{\k,~\l,~\m}$
are each expanded to second order in the deviations from the
point ($\k_0,~\l_0,~\m_0$).
After a considerable amount of algebra and a series of
transformations we find
\st
\bea\label{mm1}
i\partial_t\psi&=&\mu_0\psi + i \mmu_1\cdot\nab \psi
- \frac{1}{2} \sum_{i,j}(\mmu_2)_{ij}\partial^2_{x_ix_j}\psi\no\\
&&+~ \mu_3 \psi^+ \psi\psi  + i \mmu_4 \cdot \psi^+ \psi \nab\psi
+ i \mmu_5 \cdot \psi^+ (\nab \psi)\psi
+ i \mmu_6 \cdot (\nab \psi^+ )\psi\psi \no\\
&&+~ {\rm higher~ order~ terms},
\eea
where
\st
\be
\psi(\r,t) = \Omega^{-1/2} \sum_\etas \exp(-i \eeta\cdot\r)
a_\etas (t),
\ee
where $\Omega$ is the volume over which the members of the plane wave basis are
normalized \cite{TD89a}.
Here $\mu_i$ or $\mmu_i$ are constant parameters, determined by
matrix elements $\omega_\k$ and
$\Delta_{\k,~\l,~\m}$ and their derivatives calculated at point
($\k_0,~\l_0,~\m_0$).
To convert Eq. (\ref{mm1}) to a PDE in a complex number (c-number) field,
rather than an operator,
the center of expansion ($\k_0,~\l_0,~\m_0$) is selected to be a
critical or fixed point of the system.
The reason for this is that close to a critical point it is an
excellent approximation to replace the full quantum field, $\psi(r,t)$,
by a classical component, $\psi_c$ \cite{Ma76,Jac77,Ami78}:
\st
\be
\psi(\r,t)=\psi_c(\r,t) \hat{\I} + \hat{\phi}(\r,t),
\ee
where $\hat{\I}$ is the unit operator in Fock space,
$\psi_c$ is a c-number field, $\hat{\phi}$ is a quantum
mechanical operator with magnitude about
$|\hat{\phi}|\sim \hbar |\psi_c|$ \cite{DT95}
(see \cite{TSB97} for details).

In the next section we apply the above 
method to study the dynamic instability
of an individual microtubule.

\section{A quantum mechanical picture of the microtubule assembly processes }
\subsection{Particle states}
\noindent
Consider an individual microtubule in a free tubulin solution
containing a large number of GTP-tubulin, GDP-tubulin and a pool of free GTP
molecules.  In this solution several processes take place (as well as their reverse reactions):\\
\noindent
(i) GTP hydrolysis:
\st
\be
 \GTP \longrightarrow \GDP + \Delta_1.
\ee
(ii) generating tubulin GDP from tubulin GTP:
\st
\be
 \Tu_\GTP  ~\longrightarrow~ \Tu_\GDP + \Delta_2 ,
\ee
(iii) growth of a MT:
\st
\be
\Delta_3 + \MT_{N-1} + \Tu_\GTP ~\longrightarrow~ \MT_N,
\ee
(iv) shrinkage of a MT:
\st
\be
\MT_N  ~\longrightarrow~  \MT_{N-1} + \Tu_\GDP + \Delta_4.
\ee
Note that experimental studies determined the values of the free energies for these
reactions as:
$\Delta_1 \simeq 220$ meV, $\Delta_2 \simeq 160$ meV and
$\Delta_3 \simeq \Delta_4 \simeq 40$ meV, respectively \cite{Cap94}.  These free energies are clearly above
the thermal energy at room temperature ($k_{\rm B}T\simeq 26$ meV) and
they are within a quantum mechanical energy range that corresponds to the creation of
one or a few chemical bonds.
Hence we may consider each chemical reaction as a
quantum mechanical process \cite{GTP}. As a result, an individual
microtubule with length $L$ can be viewed as consisting of $N$
tubulin layers defining its quantum state $|N\rangle$.
A tubulin layer consists of at least one tubulin dimer and at most 13 tubulin dimers as observed in the MT's structure.
The state can
be raised/lowered by a creation/annihilation operator (i.e.
polymerization/depolymerization process) to the
$|N+1\rangle/|N-1\rangle$ state.  The corresponding MT is then
longer/shorter by one tubulin layer compared to the original one.

In this paper, in order to simplify the problem we combine the above processes into two
fundamental reactions:\\
(i) growth of a MT by one dimer by adding of one tubulin layer in
an endothermic process:
\st
\be
\Delta + \MT_{N-1} + \Tu_\GTP ~\longrightarrow~ \MT_N,
\ee
(ii) shrinkage of a MT by one dimer due to the removal of one layer of $\Tu_\GDP$ dimer in
an exothermic process:
\st
\be
\MT_N  ~\longrightarrow~  \MT_{N-1} + \Tu_\GDP + \Delta,
\ee
where $\Delta$ is the energy of the reaction.
In order to derive a quantum mechanical description of mechanisms (i) and (ii), we first
need to introduce quantum states of MT, tubulin and heat bath:
\begin{itemize}
\item $|N\rangle$ is the state of a microtubule with $N$ dimers (both GTP and GDP
tubulins).
\item $|N_T\rangle$ is the state of a tubulin, $\Tu_\GTP$ or $\Tu_\GDP$.
\item $|\tilde{N}\rangle$ is the GTP hydrolysis energy state.
\end{itemize}
Then, the relevant second quantization operators would be:
\bea
\st\label{a_dagger}
&&a^\dagger=|N+1\rangle\langle N|,\\
\st
&&a=|N-1\rangle\langle N|,\\
\st
&&b^\dagger=|N_T+1\rangle\langle N_T|,\\
\st
&&b=|N_T-1\rangle\langle N_T|,\\
\st
&&d^\dagger=|\tilde{N}+1\rangle\langle \tilde{N}|,\\
\st\label{d}
&&d=|\tilde{N}-1\rangle\langle \tilde{N}|,
\eea
Here $b$/$b^\dagger$ and $d$/$d^\dagger$ are annihilation/creation operators
of tubulin and energy quanta, respectively.  The operators $a$/$a^\dagger$ are
lowering/raising the number of tubulin layers that constructed a MT.
Following \cite{TD01}, one can express the above processes using creation and
annihilation operators (\ref{a_dagger})-(\ref{d}):
\bea
\st\label{met1}
a^\dagger b ~d  &:& \Delta + \MT_{N-1} + \Tu_\GTP \longrightarrow  \MT_N \\
\st\label{met2}
d^\dagger ~ b^\dagger ~ a &:& \hspace{2.8cm}\MT_N  \longrightarrow
\MT_{N-1} +  \Tu_\GDP + \Delta
\eea
Operators (\ref{met1}) and (\ref{met2}) describe a MT's growth and shrinkage
by one layer, respectively.  Realistically, the polymerization or depolymerization
process may happen repeatedly before reversing the process.  This can be extended within
our model by constructing product operators, i.e. $(a^\dagger b ~d)^m$ and
$(d^\dagger ~ b^\dagger ~ a)^n$, where $m$ and $n$ are the number of growing or shrinking events in
a sequence, respectively.

\subsection{The Hamiltonian}
\noindent
Based on the mechanisms in (\ref{met1}) and (\ref{met2}), the Hamiltonian
for interacting microtubules with $\Tu_\GTP/\Tu_\GDP$ tubulins can be written as
\st
\bea\label{Ham}
H&=& \sum_\k \hbar\omega_\k a_\k^\dagger a_\k
+  \sum_\m \hbar\varpi_\m b_\m^\dagger b_\m
+  \sum_\l \hbar\sigma_\l~ d_\l^\dagger d_\l\no\\
&& \hspace{2cm} +\sum_{\k,\m} \hbar( \Delta_{\k,\m} ~a^\dagger_\k b_\m ~d_{\k-\m}
+   \Delta^*_{\k,\m} ~ d^\dagger_{\k-\m}  ~b^\dagger_\m ~ a_\k),
\eea
where $\omega$, $\varpi$, $\tilde{\Delta}$ and $\Delta$ are constants in units of energy.
However, an intermediate transition between a microtubule in a growing
phase and a microtubule in a shrinking phase must also be taken into account.
A growing/shrinking microtubule may change its state
quickly or after several steps to a depolymerizing/polymerizing state and then
may change back to polymerizing/depolymerizing state.
Experimentally, the transition of microtubules from the growing to the shrinking phase
is quantified by the catastrophe rate $f_{\rm cat}$ and the transition from the shrinking to
the growing phase is expressed by
the rescue rate $f_{\rm res}$ in which $f_{\rm res} < f_{\rm cat}$.   As we discussed earlier,
these transitions can be represented
by a combination of creation and annihilation operators as the $n^{\rm th}$ power of the reaction in
(\ref{met1}) and (\ref{met2}):
\st
\bea\label{Ham1}
H&=& \sum_\k \hbar\omega_\k a_\k^\dagger a_\k
+  \sum_\m \hbar\varpi_\m b_\m^\dagger b_\m
+  \sum_\l \hbar\sigma_\l~ d_\l^\dagger d_\l \no\\
&&\hspace{2cm}+ \sum_{n=1}^\infty \sum_{\tilde{\k}_n, \tilde{\m}_n, \tilde{\l}_{n-1}}
  \hbar[ \Delta_{\tilde{\k}_n \tilde{\m}_n \tilde{\l}_n} ~ c_{\tilde{\k}_n
  \tilde{\m}_n \tilde{\l}_n}
+   \Delta^*_{\tilde{\k}_n \tilde{\m}_n \tilde{\l}_n } ~
c^\dagger_{\tilde{\k}_n \tilde{\m}_n \tilde{\l}_n} ],
\eea
where
\st
\be\label{c-op}
c_{\tilde{\k}_n \tilde{\m}_n } =
(a^\dagger_{\k_1} b_{\m_1} ~d_{\l_1}) (a^\dagger_{\k_2} b_{\m_2} ~d_{\l_2})
\ldots
(a^\dagger_{\k_n} b_{\m_n} ~d_{\l_n }).
\ee
Here $\tilde{\k}_n=\{\k_1, \k_2, \ldots, \k_n\}$ is a collection of indices and
$\sum_{\tilde{\k}_n}=\sum_{\k_1} \sum_{\k_2} \ldots \sum_{\k_n}$.
We note that the momentum conservation for the last two terms in the
Hamiltonian (\ref{Ham1}) requires that
\st
\be\label{ln}
\l_n = \sum_{i=1}^n \k_i - \sum_{i=1}^n \m_i - \sum_{i=1}^{n-1} \l_i.
\ee
Therefore, the first $n-1$ of $\l$ will be free and summed in the Hamiltonian (\ref{Ham1}).

In Bose-Einstein statistics the creation and annihilation operators satisfy
\st
\be
[ q_\k, q_\m^\dagger]=\delta_{\k \m},~~{\rm and}~~
[q_\k^\dagger,q_\m^\dagger]=0=[q_\k,q_\m],
\ee
where $[A,B]=AB-BA$ is the Dirac commutator and $q=a,b,~{\rm and}~ d$ .  Since these
operators mutually commute,
the $c_{\tilde{\k}_n \tilde{\m}_n \tilde{\l}_n}$,  Eq. (\ref{c-op}), can be rewritten as
\st
\be\label{c-op1}
c_{\tilde{\k}_n \tilde{\m}_n \tilde{\l}_n} =
a^\dagger_{\k_1} a^\dagger_{\k_2}\ldots a^\dagger_{\k_n}  b_{\m_1} b_{\m_2} ~  \ldots
b_{\m_n} ~ d_{\l_1 } d_{\l_2} \ldots d_{\l_n} =  a^\dagger_{\tilde{\k}_n}
b_{\tilde{\m}_n} d_{\tilde{\l}_n},
\ee
where $\l_n$ is given by Eq. (\ref{ln}).

\section{Classical equations of motion}
\noindent
A system of coupled equations that describes the quantum dynamics of a MT
is derived from the Heisenberg equation in Appendix A.   However, MTs are overall classical
objects (although some of their degrees of freedom may behave as quantum
observable).  As a result,  we need to ensemble average over all possible
states to obtain effective dynamical equations.

Fourier transforming of $a_\etas$, $b_\etas$ and $d_\etas$ operators
over all states, one can find
\bea
\st
&&\psi(\r,t) = \Omega^{-1/2} \sum_\etas \exp(-i \eeta\cdot\r) a_\etas (t),\\
\st
&&\chi(\r,t) = \Omega^{-1/2} \sum_\etas \exp(-i \eeta\cdot\r) b_\etas (t),\\
\st
&&\phi(\r,t) = \Omega^{-1/2} \sum_\etas \exp(-i \eeta\cdot\r) d_\etas (t),
\eea
where $\Omega$ is the volume over which the members of the plane wave basis are
normalized \cite{TD89a}.   Here $\psi(\r,t)$, $\chi(\r,t)$, and $\phi(\r,t)$
are the corresponding field operators for the quantum operators $a_\etas$, $b_\etas$ and
$d_\etas$, respectively.   The derivation of the equation of motion for the field
operators is given in Appendix B.
The final form of the equations of motion is found to be
\st
\bea\label{eq_MT0}
&& i\partial_t{\psi} = A_0 \psi + i\A_1\cdot \nab\psi
-\frac{1}{2}A_2 \nabla^2 \psi + \sum_{n=2}^\infty ~ (A_3^{(n)}\psi^*) ~{\psi^*}^{n-2}\chi^n\phi^n, \\
\st\label{eq_chi0}
&& i\partial_t{\chi} = B_0 \chi +  i\B_1\cdot \nab\chi -\frac{1}{2}B_2\nabla^2 \chi
+ \sum_{n=1}^\infty  (B_3^{(n)} \psi) \psi^{n-1}{\chi^*}^{n-1}{\phi^*}^n , \\
\st\label{eq_phi0}
&&i\partial_t{\phi} = C_0 \phi +   i\C_1\cdot \nab\phi -\frac{1}{2}C_2\nabla^2 \phi
+ \sum_{n=1}^\infty (C_3^{(n)} \psi)\psi^{n-1}{\chi^*}^n\phip^{n-1},
\eea
where $n$ represents the degree of nonlinearity corresponding to the order of chemical process underlying the term in the
equation and  $A_i$, $B_i$ and $C_i$ are constants which are given in Appendix.
We obtain the general equations of motion for the system in terms of coupled nonlinear
partial
differential equations (PDE's) that describe the MT field, the tubulin field and GTP
field, respectively.  These fields are complex function of space and time and their modulus squared corresponds to
the spatio-temporal concentration of each of the three chemical species.

\noindent
In this paper we are primarily interested in the dynamics of MTs.
Following Eq. (\ref{eq_MT0}), the dynamical equations for
growing and shrinking states of a MT up to $n=3$ can be written as
\bea
\st\label{eqMT_s}
&& i\partial_t{\psi} = A_0 \psi + i\A_1\cdot \nab\psi
-\frac{1}{2}A_2\nabla^2 \psi + (A_3^{(2)} \psi^* )~\chi^2\phi^2 + (A_3^{(3)}\psi^* )~\psi^*\chi^3\phi^3.
\eea
Furthermore, the dynamics of the tubulin field, $\chi$, and energy of the system, $\phi$, are
also determined by
\bea
\st\label{eq_chi}
i\partial_t{\chi} &=& B_0 \chi +  i\B_1\cdot \nab\chi -\frac{1}{2} B_2\nabla^2 \chi + (B_3^{(1)}\psi ) \phi^* , \\
\st\label{eq_phi}
i\partial_t{\phi} &=& C_0 \phi +  i\C_1\cdot \nab\phi - \frac{1}{2}C_2\nabla^2 \phi + (C_3^{(1)}\psi ) \chi^*,
\eea
where, for simplicity, we just keep the $n=1$ term.

Although the system of equations (\ref{eqMT_s})-(\ref{eq_phi})
is generally similar to the phenomenological system of equations
(\ref{grow})-(\ref{c_D}) found by other investigators, there are some manifest differences between them.  Equations
(\ref{eqMT_s})-(\ref{eq_phi}) are truly 3-dimensional that predict the most possible (or stable) 3D structure for a MT which is a vortex
filament (see Sec. \ref{Dynamic}) as observed experimentally.   Other observed 3D
structures of MTs such as double-wall MT, ring-shaped, sheet-like,
C-shaped and S-shaped ribbons, and hoop structures
(see \cite{Ung90,Beh06,Hab06} for details) during tubulin nucleation can be also explained through the model described here.
Furthermore, the complex nonlinearity of MTs' behavior is embedded in our system through
the role of fluctuations such as thermal noise that can generate catastrophe (see Sect.
\ref{ana_res}).

As an example, the catastrophe
event that is usually inserted by hand in those phenomenological models
is a direct result of MT's fluctuations in our model (see Eq. \ref{L4}).

A vast array of mathematical methods of finding
solutions to Eqs. (\ref{eqMT_s})-(\ref{eq_phi}) can be found in the monograph by Dixon \emph{et al.} \cite{Dix97}.
Among the analytical solutions known one can expect to find localized (solitonic) and extended (traveling wave)
solutions.
The latter ones may have the meaning of coherent oscillations that have been observed experimentally
for high tubulin concentrations by Mandelkow \emph{et al.} \cite{Man89}.   The localized solutions may correspond to
nucleation from a seed.  Sept \emph{et al.} \cite{Sep99} studied the kinetics of a set of chemical reactions
occurring during MT polymerization and depolymerization processes and also found oscillatory solutions with damping in a longer
time regime of the assembled tubulins.

\section{The dynamical equation}\label{Dynamic}

In order for the MT's assembly process to go through, one would expect that the energy in the system should
be distributed uniformly during the course of experiment.
This is based on the fact that one would expect the overall energy during the polymerization/depolymerization process to remain
conserved.  This, in fact, has been proved experimentally by direct measurements that the enthalpy change during tubulin
polymerization is about $0\pm 1$ kcal/mol of tubulin dimer \cite{Sut76}.  It is interesting to note that in some experiments,
the GTP-tubulin concentration in the solution is maintained to be constant, i.e.
$\dot{\chi}=0$ and $\nab\chi={\bf 0}$.
The uniform energy distribution requires that the energy distribution function satisfies $\dot{\phi}=0$ and $\nab\phi={\bf 0}$.
As a result of this assumption, Eq. (\ref{eq_phi}) can be solved for $\phi$ as
\st
\be\label{phi1}
\phi \sim  - q ~\psi ~\chi^*,
\ee
where $q$ is a constant parameter.  Here and thereafter, we ignore the spatial derivative term in coefficients $A_i,~B_i$
and $C_i$  to avoid further complexities.  Inserting Eq. (\ref{phi1}) into Eqs. (\ref{eqMT_s}) and (\ref{eq_chi}) we find
\st
\bea\label{eqMT_s1}
&& i\partial_t{\psi} +  i\v\cdot \nab\psi =
-\frac{1}{2}b \nabla^2 \psi +  V \psi, \\
\st\label{eq_chi1}
&&i\partial_t{\chi} =
f \nabla^2 \chi + U \chi,\\
&& V(|\psi|,|\chi|) = a + c |\chi|^4 |\psi|^2 -d |\chi|^6 |\psi|^4,\no\\
&& U(|\psi|) = e - h |\psi|^2, \no
\eea
where $|\psi|^2=\psi\psi^*$.
In the above equations we introduced a set of new parameters for simplicity.  We also
transformed $\r$ to $\r+\B_1 t$ and then set $\A_1-\B_1 =\v$ where $\v$ can be considered to represent a MT's velocity
relative to the tubulin concentration in the solution.
Here, parameters $a, b, e$ and $f$ are real but $c, d$ and $h$ are complex.   Eq. (\ref{eqMT_s1}) represents the
nonlinear cubic-quintic Schr\"{o}dinger (NLS) equation with a complex potential
that has been extensively studied in connection with topics such as
pattern formation, nonlinear optics, Bose-Einstein condensation, superfluidity
and superconductivity, etc.  \cite{AK02}.
In a series of papers, Gagnon and Winternitz discussed symmetry groups of the NLS equation and provided some exact solutions in spherical and cylindrical coordinates \cite{GW88,GW89,GW89a,GW89b,GW89c} which is of relevance
to the present case.   General solution of the NLS equation can be cast in the form of
$\psi(\r,t)=R(\r,t)\exp[iS(\r,t)]$ which involves topological defects (point in 2D and line in 3D).
In three dimensions these defects represent one-dimensional strings or vortex filaments \cite{AK02} .
In a cylindrical coordinate system, there exists a stationary solution that represents a straight vortex filament with twist:
\st
\be\label{sol}
\psi(r,\theta,z,t) = R(r) \exp[i(\omega t + n \theta + w(r) + k_z z)],
\ee
where $\omega$ is the spiral frequency,  $R(r)$ is the amplitude,
$w(r)$ is the spiral phase function and integer $n$ is the winding number of the vortex \cite{AB97}.
The axial wave number $k_z$ characterizes the vortex's twist.  $k_z=0$
represents an untwisted vortex that is the most stable solution \cite{AK02}.
In the case of the NLS equation, a family of vortices that move with a constant velocity is also a solution \cite{AB97}.

\subsection{Numerical results}
Equations (\ref{eqMT_s1}) and (\ref{eq_chi1}) are solved numerically with a no-flux boundary condition.
As an initial condition we chose a straight vortex filament perturbed by small noise (eg.
thermal or environmental noise).
In Fig. {\ref{fig:MT}
we compare the observed data on the MT length as a function of time with our simulation results.
The length of a vortex is defined as
\cite{AB97}:
\st
\be\label{L0}
L(t) = \int \Theta(\psi_0 - |\psi(\r, t)|) d^3r,
\ee
where $\Theta(x)$ is the step function and $\psi_0$ is a constant.
In Figs. {\ref{fig:MT}} and
\ref{fig:MT1} we compare the observed data on the MT length as a function of time with our simulation results.
Experimental panels in Figs. \ref{fig:MT} and \ref{fig:MT1}
represent the experimental data published by Rezania \emph{et al.} \cite{Rez08}.
Simulation panels show the numerical results of the normalized vortex length as a
function of time for the given set of
parameters. 

To provide a simple yet accurate and powerful comparison between experimental
and simulation results,
we graph recursive maps for the data points. The advantage of the recursive maps is the introduction of regularity into the data sets that allows for a better choice of adjustable parameters due to noise reduction inherent in the separation of data into subsets corresponding to independent processes.  In spite of being very simple, recursive maps of assembly and disassembly processes of individual MTs can successfully reproduce many of the key characteristic features. Consider first the following stochastic map as the simplest case that illustrates the approach taken:
\st
\be
\ell (t_n + 1) =  r[ \ell(t_n) + \delta],
\ee
where  $\ell(t_n)$ is the length of a microtubule after $n$  time steps, $t_n$.
The parameter $r$  is chosen to be a random number with the following two possibilities:
$$
r =  \left\{ \begin{array}{l}
      1 \;\;\; {\rm with \;probability}\;\; p\\
      0 \;\;\; {\rm with\; probability} \;\;1-p
      \end{array}
      \right.
      $$
In terms of the MT polymerization process, $p$ is the probability that a given event will result in assembly while $1-p$
is the probability of a complete catastrophe of the MT structure.
The above simplified model, therefore, is governed by only two
adjustable parameters: (i) the probability of a complete catastrophe
$1-p$ which is constant and independent of the length or time elapsed
and (ii) the rate of polymerization which is proportional to the length
increment $\delta$ over the unit of time chosen in the simulation.
Thus, the coefficient $\delta$ divided by the time step $\Delta t$ ($ = t_{n+1} - t_n$)
gives the average growth velocity of an individual MT.
Such information can be used to fine tune the simulation parameters.
We note that the slope of the line in the simulation panels in Fig. \ref{fig:MT} can be adjusted by varying the real parts
of parameters $c$ and $d$.  The frequency of catastrophe events can also be changed by
adjusting the parameter $b$.
In the recursive map panels in Figs. \ref{fig:MT} and \ref{fig:MT1}
we compare the recursive map for both the experimental data and the simulation results.
Based on the recursive maps, the key characteristics of the experimental and simulated
results that were obtained independently are quite similar.   This represents
Eqs. (\ref{eqMT_s1}) and truly describes the dynamics of MTs' polymerization.

To provide a more solid comparison,
a spectral analysis is also carried on both experimental and simulated data.
As discussed by Odde \emph{et al.} \cite{OBC96}, the power spectrum analysis is a more
general way to characterize the microtubule assembly/disassembly dynamics without
assuming any model \emph{a priori}.  The power spectrum panels in Figs.
\ref{fig:MT} and \ref{fig:MT1}
represent the spectral power of the experimental and simulated data, respectively.
As shown, there is a great agreement
between the experimental (solid curve) and simulated (dashed curve) spectrums.
We note that the curves are plotted at different offsets for visual clarity only.
As expected, no particular
frequency of oscillation can be found from the power spectrums.  However, both
power spectrums demonstrate a very similar broad distribution that
more or less decays with frequency as an inverse power-law with slope $\sim 1.0$ in
Fig. \ref{fig:MT} and $\sim 1.2$ in Fig. \ref{fig:MT1}, respectively.
The best fit inverse power-law is shown by a dotted line in both panels.

More interestingly, our results show no attenuation states during MT polymerization
(Fig. \ref{fig:MT}).  The MT length undergoes small fluctuations all the time. This
can be understood by noting that our model is based on the cyclic polymerization and
depolymerization of tubulin dimers.  Behavior consistent with this result has recently been observed
by Schek \emph{et al.} \cite{Sch07} who studied the microtubule assembly dynamics
at higher spatial ($\sim$ 1-5 nm) and temporal ($\sim$ 5 kHz) resolutions.
They found that even in the growth phase, a MT undergoes shortening excursions at the
nanometer scale.

\subsection{Analytical results}\label{ana_res}
Using Eq. (\ref{L0}), the time-varying nature of vortex filaments' length can also be extracted analytically.
Taking time derivative of $L(t)$ one finds
\bea\label{L1}
&&\frac{d}{dt} L(t) \sim  \int \frac{\partial \xi}{\partial t} \partial_\xi\Theta(\xi)  d^3r,\no\\
&&~~~~~~~~~=- \frac{1}{2}\int\frac{1}{|\psi|} \partial_{t}(\psi\psi^*) \delta(\psi_0-|\psi|) d^3r, \no\\
&&~~~~~~~~~=- \frac{1}{2}\int\frac{1}{|\psi|}  [-\v\cdot\nab(\psi\psi^*)
+(ib/2)(\psi^*\nabla^2\psi -\psi\nabla^2 \psi^*)-i(V-V^*)\psi\psi^*] \delta(\psi_0-|\psi|) d^3r, \no
\eea
where $\xi=\psi_0-|\psi(\r,t)|$ and $\delta(x)$ is the Dirac delta function.  For a general vortex solution, this
may lead to a very complicated function of time.
However, for the given vortex solution, Eq. (\ref{sol}), the above statement will be simplified as
\st
\bea\label{dL}
&&\frac{d}{dt} L(t) \sim  \int [\v\cdot\nab R+ (b/4) (4\nab R\cdot\nab w + 2 R\nabla^2w)+Im(V) R] \delta(\psi_0-|\psi|) d^3r,
\eea
where the above integral is constant in time.  As a result, the vortex length can grow in time linearly.
In order to calculate possible fluctuations in the vortex length,
we introduce a small perturbation (due the thermal or environmental noise)
$\delta \psi(\r,t) \sim \xi(\r) \exp(i\omega' t)$ where $|\xi| \ll 1$ and $\omega'$ represents
the quasi-periodicity of the fluctuations.
Inserting it into Eq. (\ref{L0}) and taking the time derivative, we have
\st
\bea\label{L2}
&&\frac{d}{dt}(L+\delta L)= \int \partial_t\Theta(\psi_0 - |\psi+\delta\psi|) d^3r,\no\\
&&~~~~~~~~~~~~~~\approx-\frac{1}{2}\int\frac{1}{|\psi+\delta\psi|} \partial_{t}[\psi\psi^* +\psi^*\delta\psi
+\psi\delta\psi^*  ]\delta(\psi_0-|\psi|) d^3r.~~~
\eea
Taking the time derivative and expanding the denominator in the integral we have
\bea
&&\partial_{t}(\psi\psi^*) = -\v\cdot\nab(\psi\psi^*) + \frac{ib}{2}[\psi^*\nabla^2\psi -  \psi\nabla^2 \psi^*-i(V-V^*)\psi\psi^*],\no\\
&&\partial_{t}(\psi\delta\psi^*+\psi^*\delta\psi)=  -\v\cdot\nab\psi ~\delta\psi^*
+\frac{ib}{2}\delta\psi^*\nabla^2\psi -i(\omega'+V)\psi\delta\psi^*+  c.c. ~,\no\\
&&|\psi+\delta\psi|^{-1}\approx \frac{1}{|\psi|}\left[1-\frac{1}{2}\frac{1}{|\psi|^2}\psi\delta\psi^*
-\frac{1}{2}\frac{1}{|\psi|^2}\psi^*\delta\psi\right],  \no
\eea
where $c.c.$ means the complex conjugate.
Again, for the given solution (\ref{sol}) one finds after some manipulations
\st
\be\label{L3}
\delta L\approx \frac{\Gamma}{\omega'-\omega}\sin(\omega-\omega')t 
+{\rm const.},
\ee
where
\be
\Gamma=\int[-\v\cdot\nab F+(ib/2)\nabla^2 F -i(V+\omega') F -
(1/2) F/|F|^3]\xi^*~\delta(\psi_0-|\psi|)~ d^3r + c.c. ,\no
\ee
and $F(\r)=R(\r)\exp[i(w(r)+n\theta+k_z z)]$.  Interestingly, from Eq. (\ref{L3}) one can see that even if
$\omega'$ goes to zero, the vortex length fluctuates with the spiral frequency $\omega$.  As a result,
the vortex length fluctuates all the time.   Furthermore, when $\omega' \rightarrow \omega$ Eq. (\ref{L3}) reduces to
\st
\be\label{L4}
\delta L\approx - \Gamma t 
+{\rm const.},
\ee
where represents a linear shortening similar to catastrophe events.
As a result, a catastrophe event can be
explained within our model.


\section{Discussion}
\noindent
In our model, the basic structural unit is the tubulin dimer. Each dimer exists in a quantum
mechanical state characterized by several variables even in our simplified approach.
Each microstate of a tubulin dimer  is sensitive to the states of its neighbors.
Tubulin dimers have both discrete degrees of freedom (distribution of charge) and
continuous degrees of freedom (orientation). A model that focuses on the discrete
ones will be an array of coupled binary switches \cite{Cam01,Ras90},
while a model that focuses on the continuous ones will probably be an array of coupled
oscillators \cite{BT97,Sam92}.  In the present paper
we have focused on tubulin binding and GTP hydrolysis as the key processes determining
the states of microtubules. These are also the degrees of freedom that are most easily
accessible to experimental determination.  In this paper we have shown how a quantum
mechanical description of the energy binding reactions taking place during MT polymerization
can lead to nonlinear field dynamics with very rich behavior that includes both localized
energy transfer and oscillatory solutions.

In particular, based on the chemical binding reactions that are taking place during microtubule
polymerization, a quantum mechanical Hamiltonian for the system is proposed.
Equations of motion are then derived and transformed from the purely quantum mechanical
description to a semi-classical picture using the inverse Fourier  transformation. 
After lengthy calculations we found that the dynamics of a MT can be explained by the cubic-quintic nonlinear
Schr\"{o}dinger equation (NLS) with variable coefficients.  A generic solution of the NLS equation in cylindrical
geometry is a vortex filament \cite{AK02,GW89c,AB97}.
Interestingly, we showed both analytically and numerically that
such a solution can grow or shrink linearly in time as well as fluctuate temporally with some frequency.
This behavior exhibits two distinct dynamical phases: (a) linear
growth/shrinkage and (b) oscillation about some mean value, and is consistent with the characteristics of
the MT's dynamics as observed in different controlled experiments in vitro (Fig. 1).

It is noteworthy that dynamics of pattern formation can be also described by NLS equation in which
$\psi(\r,t)$ represents the order parameter.  Interestingly, a number of convincing experiments,
performed by Tabony \emph{et al.} \cite{Tab}
demonstrated that gravity can indeed influence certain chemical reactions.
Tabony and his colleagues, at the French Atomic Energy Commission lab in Grenoble,
found that when cold solutions of purified tubulin and the energy-releasing compound GTP
were warmed to body temperature, microtubules formed in distinct bands. These bands form
at right angles to the orientation of the gravity field or, if spun, to the centrifugal force.
Despite several studies \cite{Por03,Por05,AT05,AT06}, the above experiments are yet to be fully explained theoretically.
Our goal in future studies is to
focus on the dynamics of pattern formation by MTs using the results presented in this paper.

We have demonstrated here that the
assembly process can be described using quantum mechanical principles applied to biochemical
reactions.   This can be subsequently transformed into a highly nonlinear
semi-classical dynamics problem.  The gross features of MT dynamics satisfy
classical field equations in a coarse-grained picture.  Individual chemical
reactions involving the constituent molecules still retain their quantum
character.  The Fourier transformation 
allows for a simultaneous
classical representation of the field variables and a quantum approach to their
fluctuations.   Here, the overall MT structure (and their ensembles) can be
viewed as a virtual classical object in (3+1) dimensional space-time.   However,
at the fundamental level of its constituent biomolecules, it is quantized as
are true chemical reactions involving its assembly or disassembly.

{\bf Acknowledgment}\\
This research was supported in part by the Natural Sciences and
Engineering Research Council of Canada (NSERC) and the Canadian Space Agency (CSA).
Insightful discussions with S. R. Hameroff and  J. M. Dixon are gratefully
acknowledged.  The authors would also thank  L. Wilson for providing microtubule assembly data.
VR specially thanks I. Aranson for sharing his CGLE code and for fruitful discussions.

\setcounter{sub}{0}
\setcounter{subeqn}{0}
\renewcommand{\theequation}{A\thesub\thesubeqn}

\appendix
\section{Derivation of the Heisenberg equations of motion}

\noindent
The Heisenberg equation of motion for a space- and time-dependent operator $q(\r,t)$
reads as
\st
\be
i\hbar \partial_t{q}(\r,t) = - [H,q(\r,t)],
\ee
where $H$ is the Hamiltonian.  Before finding equations
of motion, one needs
to calculate the commutation relation $[ q_\etas, q^\dagger_{\tilde{\k}_n}]$ that is
\st
\bea\label{qq0}
&&[ q_\etas, q^\dagger_{\tilde{\k}_n}]=[q_\etas, q^\dagger_{\k_1} \ldots
q^\dagger_{\k_n}] =
 \delta_{\etas,  \k_1}~ q^\dagger_{\k_2} q^\dagger_{\k_3} \ldots q^\dagger_{\k_n  }
+  \delta_{\etas,  \k_2}~ q^\dagger_{\k_1} q^\dagger_{\k_3} \ldots q^\dagger_{\k_n  }
+ \ldots \no\\
&&\hspace{6cm}+
 \delta_{\etas,  \k_n}~  q^\dagger_{\k_1} q^\dagger_{\k_2} \ldots q^\dagger_{\k_{n-1}}.
\eea
Since all $\k_1, \k_2, \ldots, \k_n$ are dummy indices one can write Eq. (\ref{qq0}) as
\st
\be\label{qq}
[ q_\etas, q^\dagger_{\tilde{\k}_n}]= n~\delta_{\etas, \k_n} ~q^\dagger_{\tilde{\k}_{n-1}},
\ee
where $\k_n$ is chosen for simplicity.
Using Eq. (\ref{qq}) we can find the commutation relations between
$a_\etas$ and $b_\etas$
operators with $c_{\tilde{\k}_n \tilde{\m}_n \tilde{\l}_n}$ and
$c_{\tilde{\k}_n \tilde{\m}_n \tilde{\l}_n}^\dagger$ operators as
\st
\bea
&&[ a_\etas, c_{\tilde{\k}_n \tilde{\m}_n \tilde{\l}_n}  ]
= [ a_\etas, a^\dagger_{\tilde{\k}_n} b_{\tilde{\m}_n} d_{\tilde{\l}_n } ]
= n~\delta_{ \etas, \k_n  } a^\dagger_{\tilde{\k}_{n-1}} b_{\tilde{\m}_n} d_{\tilde{\l}_n}, \\
\st
&&[ b_\etas, c^\dagger_{\tilde{\k}_n \tilde{\m}_n \tilde{\l}_n } ]
= [ b_\etas,   d^\dagger_{\tilde{\l}_n } b^\dagger_{\tilde{\m}_n}
a_{\tilde{\k}_n} ]
= n~\delta_{ \etas, \m_n } d^\dagger_{\tilde{\l}_n }
b^\dagger_{\tilde{\m}_{n-1}} a_{\tilde{\k}_n}.
\eea
However, the commutation relation between $d_\etas$ operator and
$c_{\tilde{\k}_n \tilde{\m}_n \tilde{\l}_n}^\dagger$
will be
\st
\be
[ d_\etas, c^\dagger_{\tilde{\k}_n \tilde{\m}_n \tilde{\l}_n } ]
= [ d_\etas,   d^\dagger_{\tilde{\l}_n } b^\dagger_{\tilde{\m}_n}
a_{\tilde{\k}_n} ]
= \left( (n-1)~ \delta_{ \etas, \l_{n-1} } d^\dagger_{\tilde{\l}_{n-2} }  d^\dagger_{\l_n}
+ ~ \delta_{ \etas, \l_{n} } d^\dagger_{\tilde{\l}_{n-1} }  \right)
 b^\dagger_{\tilde{\m}_n}  a_{\tilde{\k}_n},
\ee
where $\l_n$ is given by Eq. (\ref{ln}).
Therefore, the equation of motion for
$a_\etas$, $b_\etas$ and $d_\etas$ operators (and their Hermitian conjugates)
can be derived from Hamiltonian (\ref{Ham1}) as
\bea
\st\label{eq1}
&&i\partial_t{a}_\etas = \omega_\etas a_\etas
          +  \sum_n  \sum_{\tilde{\k}_{n-1} \tilde{\m}_n \tilde{\l}_{n-1}}
  n~ \Delta_{\etas \tilde{\k}_{n-1} \tilde{\m}_n \tilde{\l}_{n-1}}
  a^\dagger_{\tilde{\k}_{n-1}} b_{\tilde{\m}_{n}}  d_{\tilde{\l}_{n-1}}
d_{\etas+\sum_{i=1}^{n-1} (\k_i-\l_i) - \sum_{i=1}^n \m_i  }
,~~~~~~~\\
\st\label{eq2}
&&i \partial_t{b}_\etas = \varpi_\etas b_\etas +
\sum_n  \sum_{\tilde{\k}_{n} \tilde{\m}_{n-1} \tilde{\l}_{n-1}}
  n~ \Delta_{\etas \tilde{\k}_{n} \tilde{\m}_{n-1} \tilde{\l}_{n-1}}
 ~ d^\dagger_{\tilde{\l}_{n-1} }
d^\dagger_{{\sum_{i=1}^{n} \k_i - \etas- \sum_{i=1}^{n-1}( \m_i + \l_i)} }
b^\dagger_{\tilde{\m}_{n-1}}
 a_{\tilde{\k}_{n}}
,~~~~~~~~\\
\st\label{eq3}
&&i \partial_t{d}_\etas = \sigma_\etas d_\etas
+  \sum_n  \sum_{\tilde{\k}_n \tilde{\m}_{n} \tilde{\l}_{n-2}}
(n-1)~\Delta_{\etas \tilde{\k}_n \tilde{\m}_{n}\tilde{\l}_{n-2} }
d^\dagger_{\tilde{\l}_{n-2}} d^\dagger_{{\sum_{i=1}^{n} (\k_i-\m_i) -
\etas- \sum_{i=1}^{n-2} \l_i} }
 ~b^\dagger_{\tilde{\m}_{n}}
a_{\tilde{\k}_{n}}\no\\
&&\hspace{2.5cm}
+  \sum_n  \sum_{\tilde{\k}_n \tilde{\m}_{n} \tilde{\l}_{n-1}}
\delta_{\etas, {{\sum_{i=1}^{n} (\k_i-\m_i) - \sum_{i=1}^{n-1} \l_i} }}
 ~\Delta_{\tilde{\k}_n \tilde{\m}_{n}\tilde{\l}_{n-1} }
d^\dagger_{\tilde{\l}_{n-1}}
 ~b^\dagger_{\tilde{\m}_{n}}
a_{\tilde{\k}_{n}}.
\eea
\setcounter{sub}{0}
\setcounter{subeqn}{0}
\renewcommand{\theequation}{B\thesub\thesubeqn}

\section{Derivation of equation of motion for the field operators}
\noindent
Multiplying both sides of Eq. (\ref{eq1}) by $\exp(-i\eeta\cdot\r)$, dividing by
$\Omega^{1/2}$ and
summing over $\eeta$, one finds
\st
\bea\label{eq1_a}
&&i\partial_t\psi_ = \Omega^{-1/2}\ms \sum_\etas \omega_\etas \exp(-i\eeta\cdot\r)
a_\etas\right.\no\\
&&\left. \hspace{0cm}  + \sum_n  \sum_\etas \sum_{\tilde{\k}_{n}
\tilde{\m}_n \tilde{\l}_{n-1}}
n~ \Delta_{\etas \tilde{\k}_{n-1} \tilde{\m}_n \tilde{\l}_{n-1}} \exp(-i\eeta\cdot\r)~
  a^\dagger_{\tilde{\k}_{n-1}} b_{\tilde{\m}_{n}}  d_{\tilde{\l}_{n-1}}
d_{\etas+\sum_{j=1}^{n-1} (\k_j-\l_j) - \sum_{j=1}^n \m_j  }
 \rs.\no\\
\eea
Changing $\eeta \rightarrow
\eeta - \sum_{j=1}^{n-1} (\k_j-\l_j) + \sum_{j=1}^n \m_j  $ in the
second term of Eq. (\ref{eq1_a}), one  finds
\st
\bea\label{eq1_b}
&&i \partial_t\psi_ = \Omega^{-1/2}\ms \sum_\etas \omega_\etas \exp(-i\eeta\cdot\r)
a_\etas\right.\no\\
&& \hspace{3cm}  +
\sum_n  \sum_\etas \sum_{\tilde{\k}_{n} \tilde{\m}_n \tilde{\l}_{n-1}}
n~ \Delta_{\etas-\xis~ \tilde{\k}_{n-1} \tilde{\m}_n \tilde{\l}_{n-1}}
e^{-i\etas\cdot\r + i \sum_{j=1}^{n-1} (\k_j-\l_j)\cdot \r - i \sum_{j=1}^n \m_j\cdot\r}\no\\
&&\left. \hspace{5cm} \times ~
  a^\dagger_{\tilde{\k}_{n-1}} b_{\tilde{\m}_{n}}  d_{\tilde{\l}_{n-1}}
d_\etas
\rs,
\eea
or
\st
\bea\label{eq1_c}
&&i \partial_t\psi_ = \Omega^{-1/2}\ms \sum_\etas \omega_\etas \exp(-i\eeta\cdot\r)
a_\etas + ~\sum_n \sum_\etas \sum_{\tilde{\k}_{n} \tilde{\m}_n \tilde{\l}_{n-1}}
n~ \Delta_{\etas-\xis~ \tilde{\k}_{n-1} \tilde{\m}_n \tilde{\l}_{n-1}} \right. \no\\
&& \hspace{4cm}\left. \times
~e^{-i\etas\cdot\r}~ d_\etas~
e^{i \tilde{\k}_{n-1}\cdot \r}~ a^\dagger_{\tilde{\k}_{n-1}} ~
e^{-i \tilde{\m}_{n}\cdot\r} ~b_{\tilde{\m}_{n}}  ~
e^{-i \tilde{\l}_{n-1}\cdot \r}~ d_{\tilde{\l}_{n-1}}~ \frac{}{}
\rs,
\eea
where $\xxi =  \sum_{j=1}^{n-1} (\k_j-\l_j) - \sum_{j=1}^n \m_j$.
Here, for example,
$\exp(-i\tilde{\k}_n\cdot \r) = \exp(-i\k_1\cdot \r) \exp(-i\k_2\cdot \r) \ldots
\exp(-i\k_n\cdot \r) = \exp(-i\sum_{j=1}^n \k_j\cdot \r)$.
Our goal is now to rewrite Eq. (\ref{eq1_c}) in terms of field operators,
$\psi,~ \chi,~ \phi$,
and their derivatives.  This can be done in a straightforward manner provided the
dispersion matrix elements $\omega_\etas$ and
$ \Delta_{\etas-\xis~ \tilde{\k}_{n-1} \tilde{\m}_{n} \tilde{\l}_{n-1}}$
which are generally
function $\eeta$, $\k_i$, $\m_i$ and $\l_i$
($1\leq i \leq n$) are known.  Unfortunately, such information is very model
dependent.  Therefore, the simplest
way that also keeps the generality of the problem is to Taylor expand these
matrix elements about some point
($\eeta_0,\k_{0i},\m_{0i}, \l_{0i}$) in the space spanned by $\eeta$,
$\k_i$, $\m_i$ and $\l_i$ \cite{TD89a,DT95}.

Expanding $\omega_\etas$ to all orders, one finds
\st
\be\label{omega}
\omega_\etas = \omega_0 + \sum_{s=1}^\infty
[(\eeta-\eeta_0)\cdot \nab_\etas]^s\omega_0/s!\;,
\ee
where $\omega_0 =\omega_{\etas_0}$.  Furthermore, for any function
$ f(\eeta, \tilde{\k}_n, \tilde{\m}_n, \tilde{\l}_n)= \Delta_{\etas \tilde{\k}_{n} \tilde{\m}_{n} \tilde{\l}_{n}}$
we can write
\st
\bea\label{f}
&&f(\eeta, \tilde{\k}_n, \tilde{\m}_n, \tilde{\l}_n) = f_0
+ (\eeta-\eeta_0)\cdot\nabla_\etas f|_0
+ \sum_{j=1}^n(\k_j-\k_{0j})\cdot\nab_{\k_j} f|_0
\no\\
&&\hspace{3.5cm}
+ \sum_{j=1}^n(\m_j-\m_{0j})\cdot\nabla_{\m_j} f|_0
+ \sum_{j=1}^n(\l_j-\l_{0j})\cdot\nab_{\l_j} f|_0 + \ldots,
\no\\
&& \hspace{.2cm}+ \sum_{p,q,r=1}^n \sum_{s=2}^\infty \sum_{u=0}^s
\sum_{v=0}^{u} \sum_{w=0}^{s-u-v}
 {^s}C_u~~ {^{u}}C_v ~~ {^{s-u-v}}C_w /s! ~~ ~\no\\
&& \hspace{0.1cm} \times ~
 [(\eeta-\eeta_0)\cdot \nab_\etas]^u~
[(\k_p-\k_{0p})\cdot \nab_{\k_p}]^v ~
[(\m_q-\m_{0q})\cdot \nab_{\m_q}]^w ~[(\l_r-\l_{0r})\cdot
\nab_{\l_r}]^{s-u-v-w}f|_0,\no\\
\eea
where
\st
\be
f_0 = \sum_{p, q, r=1}^n f(\eeta_0,\k_{0p}, \m_{0q}, \l_{0r}),
\ee
where $^sC_r$ are binomial coefficients.
Here, for example, $\nab_{\m} f$ means $\hat{i} \partial_{m_{x}} f  + \hat{j}
\partial_{m_{y}} f + \hat{k} \partial_{m_{z}} f$ where $\hat{i}, \hat{j}$
and $\hat{k}$ are unit vectors in the $m_x, m_y$ and $m_z$ directions,
respectively, and $\nab_\m f|_0$ is the value of the gradient at point
($\eeta_0, \k_0, \m_0, \l_0$).

Using Eqs. (\ref{omega}) and (\ref{f}), Eq. (\ref{eq1_c}) can be written as
\st
\bea\label{eq1_d}
&&i \partial_t\psi = \lambda_0(\omega) \psi + i\llambda_1(\omega)\cdot \nab\psi
- {1\over 2} \sum_{i,j} [\llambda_2(\omega)]_{ij}\partial_{x_i x_j}^2 \psi
+ \sum_n n \Omega^{(3n-1)/2} \Lambda_{1}^{(n)} \psip^{n-1} \chi^n \phi^{n}\no\\
&& + \sum_n n \Omega^{(3n-1)/2} \left( \psip^{n-1}\chi^n\phi^{n-1}\nab_\etas f|_0
\cdot \nab\phi
+ \sum_{j=1}^{n-1} \nab_{\k_j}f|_0 \cdot \nab\psip ~\psip^{n-2}\chi^n\phi^n
 \right.\no\\
&&\hspace{2cm}+\left. \sum_{j=1}^{n} \psip^{n-1}\nab_{\m_j}f|_0 \cdot \nab\chi
~\chi^{n-1}\phi^n
+ \sum_{j=1}^{n-1}\psip^{n-1}\chi^n \nab_{\l_j}f|_0 \cdot \nab\phi~ \phi^{n-1} \right),
\eea
where
\st
\bea
&&\lambda_0(\omega) = \omega_0 - \eeta_0\cdot\nab_\eta\omega|_0 + (1/2)\sum_{i,j}
  \eta_{0i}\eta_{0j}
\partial^2_{\eta_i\eta_j}\omega|_0~, \\
\st
&&[\llambda_1(\omega)]_i = -\sum_j\eta_{0j}\partial^2_{\eta_i\eta_j} \omega|_0
+ \partial_{\eta_i} \omega|_0~,\\
\st
&&[\llambda_2(\omega)]_{ij} = \partial^2_{\eta_i\eta_j} \omega|_0,\\
\st
&& \Lambda_{1}^{(n)} = f_0 -\eeta_0\cdot\nab_\eta f|_0 - \sum_{j=1}^{n-1}
\k_{0j}\cdot\nab_{k_j} f|_0
- \sum_{j=1}^{n} \m_{0j}\cdot\nab_{m_j}f|_0  - \sum_{j=1}^{n-1}
 \l_{0j}\cdot\nab_{l_j} f|_0.~~~~~~~~
\eea
Similarly, using Eqs. (\ref{eq2}) and (\ref{eq3}), one can write equations of
 motion for $\chi$ and $\phi$
as
\st
\bea\label{eq2_b}
&&i \partial_t\chi = \lambda_0(\varpi) \chi + i\llambda_1(\varpi)\cdot \nab\chi
- {1\over 2} \sum_{i,j} [\llambda_2(\varpi)]_{ij}\partial_{x_i x_j}^2 \chi
+ \sum_n n \Omega^{(3n-1)/2} \Lambda_2^{(n)} \psi^{n} \chip^{n-1} \phip^{n}\no\\
&& + \sum_n n \Omega^{(3n-1)/2} \left( \psi^{n}\chip^{n-1}\phip^{n-1}\nab_\etas
f|_0 \cdot \nab\phip
+ \sum_{j=1}^{n} \nab_{\k_j}f|_0 \cdot \nab\psi ~\psi^{n-1}\chip^n\phip^n
 \right.\no\\
&&\hspace{2cm}+\left. \sum_{j=2}^{n-1} \psi^{n}\nab_{\m_j}f|_0 \cdot \nab\chip
~\chi^{n-2}\phi^n
+ \sum_{j=1}^{n-1}\psi^{n}\chip^{n-1} \nab_{\l_j}f|_0 \cdot \nab\phip~ \phip^{n-1} \right),\\
\st\label{eq3_b}
&&i\partial_t\phi = \lambda_0(\sigma) \phi + i\llambda_1(\sigma)\cdot \nab\phi
- {1\over 2} \sum_{i,j} [\llambda_2(\sigma)]_{ij}\partial_{x_i x_j}^2 \phi
+ \sum_n n \Omega^{(3n-1)/2} \Lambda_3^{(n)} \psi^{n} \chip^n \phip^{n-1}\no\\
&& + \sum_n \Omega^{(3n-1)/2} \left( (n-1) \psi^{n}\chip^n\phip^{n-2}
\nab_\etas f|_0 \cdot \nab\phip
+  n  \sum_{j=1}^{n} \nab_{\k_j}f|_0 \cdot \nab\psi ~\psi^{n-1}\chip^n\phi^{n-1}  \right.\no\\
&&\hspace{2cm}+\left. n\sum_{j=1}^{n} \psi^{n}\nab_{\m_j}f|_0 \cdot \nab\chip
 ~\chip^{n-1}\phip^{n-1}
+ n\sum_{j=1}^{n-1}\psi^{n}\chip^n \nab_{\l_j}f|_0 \cdot \nab\phip~ \phip^{n-2} \right),\no\\
\eea
where
\st
\bea
&& \Lambda_{2}^{(n)} = f_0 -\eeta_0\cdot\nab_\eta f|_0 - \sum_{j=1}^{n}
\k_{0j}\cdot\nab_{k_j} f|_0
- \sum_{j=1}^{n-1} \m_{0j}\cdot\nab_{m_j}f|_0  - \sum_{j=1}^{n-1}
 \l_{0j}\cdot\nab_{l_j} f|_0,~~~~~~~~~\\
\st
&& \Lambda_{3}^{(n)} = f_0 -\eeta_0\cdot\nab_\eta f|_0 - \sum_{j=1}^{n}
\k_{0j}\cdot\nab_{k_j} f|_0
- \sum_{j=1}^{n} \m_{0j}\cdot\nab_{m_j}f|_0  - \sum_{j=1}^{n-1}
 \l_{0j}\cdot\nab_{l_j} f|_0.~~~~~~~~~
\eea
Simplifying the equations of motion as
\st
\bea\label{eq1_eA}
&& i \partial_t{\psi} = A_0 \psi + i\A_1\cdot \nab\psi
- \frac{1}{2}A_2 \nabla^2 \psi+ \sum_n ~( A_3^{(n)} \psip)~\psip^{n-2}\chi^n\phi^n, \\
\st\label{eq2_dA}
&& i \partial_t{\chi} = B_0 \chi + i\B_1\cdot \nab\chi - \frac{1}{2} B_2\nabla^2 \chi
+ \sum_n ( B_3^{(n)} \psi) \psi^{n-1}\chip^{n-1}\phip^n ,\\
\st\label{eq3_dA}
&&i \partial_t{\phi} = C_0 \phi +  i\C_1\cdot \nab\phi -\frac{1}{2}C_2\nabla^2 \phi
+ \sum_n  ( C_3^{(n)} \psi )  \psi^{n-1}\chip^n\phip^{n-1},
\eea
where
\st
\bea
&&A_0 = \lambda_0(\omega),~~\A_1=\llambda_1(\omega), ~~A_2=\lambda_2(\omega),~~
A_3^{(n)}\psip=  n \Omega^{3n-1\over 2}(\Lambda_1^{(n)} +  \sum_{j=1}^{n-1} \nab_{\k_j}f|_0 \cdot \nab)\psip ,~~~~\no\\
&&\\
\st
&&B_0 = \lambda_0(\varpi),~~\B_1=\llambda_1(\varpi),~~B_2=\lambda_2(\varpi),~~
B_3^{(n)}\psi=  n \Omega^{3n-1\over 2}( \Lambda_2^{(n)} + \sum_{j=1}^{n} \nab_{\k_j}f|_0 \cdot \nab)\psi  ,\no\\
&&\\
\st
&&C_0 = \lambda_0(\sigma),~~\C_1=\llambda_1(\sigma),~~C_2=\lambda_2(\sigma),~~
C_3^{(n)}\psi=  n \Omega^{3n-1\over 2} ( \Lambda_3^{(n)} + \sum_{j=1}^{n} \nab_{\k_j}f|_0 \cdot \nab)\psi.
\eea

----------------------------------------------------------------------------------------


\clearpage

\begin{flushleft}
\begin{table}[th]
\caption{The parameters available in the literature. }\label{Tab1}
\begin{tabular}{lcll}
\hline
Parameter & Simulation Coeff.~~~ & ~~~~Exp. Value~~ & Reference\\\hline
MT growth rate  & Real($c$) & $0.50 - 19.7$ ($\mu$m/min)& \cite{SW93,OCB95,CFK95,Fey95,Gild92,PW02,Arnal00,Rusan}\\
MT shortening rate   & Real($d$) &   $4.1 - 34.9  $ ($\mu$m/min)& \cite{Gild92,SW93,CFK95,Arnal00,PW02,Rusan}\\
MT catastrophe frequency  & & $0.12 - 3.636$ (/min)& \cite{SW93,CFK95,Arnal00,PW02}\\
MT diffusion constant   &  $b$ & $2.6-30.3$ ($\mu$m$^2$/min)& \cite{Vor97,Vor99,Maly02}\\
Tubulin diffusion constant ~~ & $e$ & $300-480$ ($\mu$m$^2$/min)& \cite{Dog95,Odde97} \\
\end{tabular}
\end{table}
\end{flushleft}

\clearpage
\noindent {\Large{\bf Figure Legends}}\\

\noindent {{\bf Figure~\ref{fig:MT}.}:}
Length of a distinct microtubule as a
function of time. The top-left panel represent experimental data published in \cite{Rez08}.
The top-right panel is the simulation result with the set of parameters
($a=1, b = 10, c = 10 + i, d=20 + i, e =300, f =1$ and $h= -.1+i$).  The bottom-left
panel shows recursive maps for both experimental and simulation results.
The bottom-right panel represents the power spectrum of the experimental (solid curve)
and simulated (dashed curve) data, respectively.
The curves are plotted at different offsets for clarity. No particular
frequency of oscillation can be seen from the power spectrums.  However, both
power spectrums show a very similar broad distribution that
more or less decays with frequency as an inverse power-law with slope $\sim 1.0$.
The best fit inverse
power-law is shown by a dotted line.

\noindent {{\bf Figure~\ref{fig:MT1}.}:}
Same as Fig 1. but with the set of parameters
($a=1, b = 30, c = 10 + 10i, d=20 +10 i, e =300, f =1$ and $h=
-.1+i$).  In the power spectrum panel, the inverse power-law has a slope of
$\sim 1.2$.

\clearpage

\begin{center}
\begin{figure}
\vspace{-.5cm}
\includegraphics[width=1\hsize]{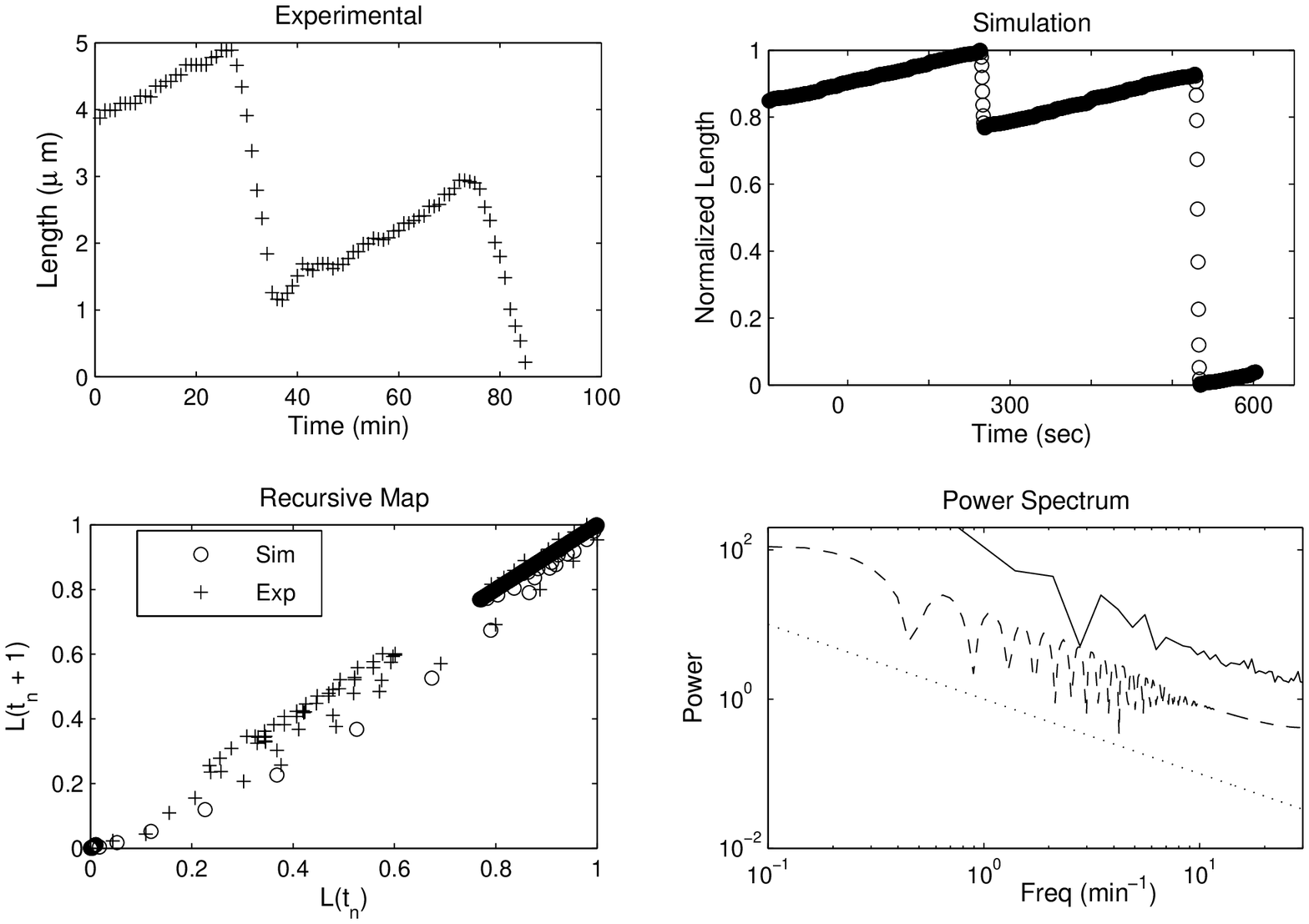}
\vspace{-.5cm} \caption{
}\label{fig:MT}
\end{figure}
\end{center}

\begin{center}
\begin{figure}
\vspace{-.5cm}
\includegraphics[width=1.\hsize]{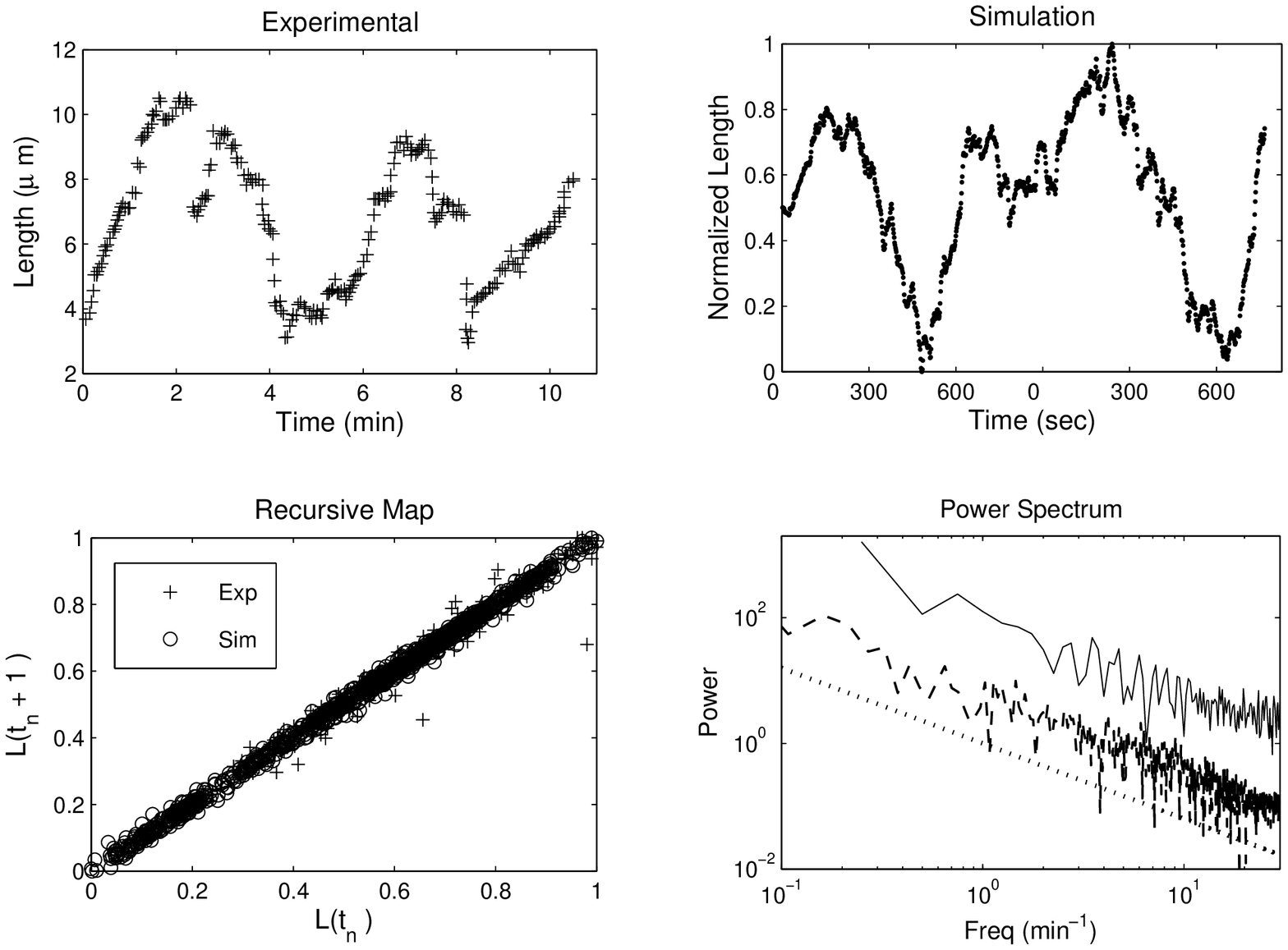}
\vspace{-.5cm} \caption{
}\label{fig:MT1}
\end{figure}
\end{center}


\begin{thebibliography}{59}
\bibitem{GTP}
We note that GTP hydrolysis is a subtle biochemical
process that carries a quantum of biological energy and thus allows us
to make a link between quantum mechanics and polymer dynamics.
\end{thebibliography}
\end{document}